\documentclass{article}
\usepackage{amsmath, amssymb, amsfonts}
\title{On the time dependent Schwarzschild - de Sitter spacetime } 
\author{Hristu Culetu, \\Ovidius University, Dept.of Physics, \\B-dul Mamaia 124, 900527 Constanta, Romania, \\e-mail : hculetu@yahoo.com}

\begin{document}
\numberwithin{equation}{section}
\pagenumbering{arabic}
\maketitle
\newcommand{\fv}{\boldsymbol{f}}
\newcommand{\tv}{\boldsymbol{t}}
\newcommand{\gv}{\boldsymbol{g}}
\newcommand{\OV}{\boldsymbol{O}}
\newcommand{\wv}{\boldsymbol{w}}
\newcommand{\WV}{\boldsymbol{W}}
\newcommand{\NV}{\boldsymbol{N}}
\newcommand{\hv}{\boldsymbol{h}}
\newcommand{\yv}{\boldsymbol{y}}
\newcommand{\RE}{\textrm{Re}}
\newcommand{\IM}{\textrm{Im}}
\newcommand{\rot}{\textrm{rot}}
\newcommand{\dv}{\boldsymbol{d}}
\newcommand{\grad}{\textrm{grad}}
\newcommand{\Tr}{\textrm{Tr}}
\newcommand{\ua}{\uparrow}
\newcommand{\da}{\downarrow}
\newcommand{\ct}{\textrm{const}}
\newcommand{\xv}{\boldsymbol{x}}
\newcommand{\mv}{\boldsymbol{m}}
\newcommand{\rv}{\boldsymbol{r}}
\newcommand{\kv}{\boldsymbol{k}}
\newcommand{\VE}{\boldsymbol{V}}
\newcommand{\sv}{\boldsymbol{s}}
\newcommand{\RV}{\boldsymbol{R}}
\newcommand{\pv}{\boldsymbol{p}}
\newcommand{\PV}{\boldsymbol{P}}
\newcommand{\EV}{\boldsymbol{E}}
\newcommand{\DV}{\boldsymbol{D}}
\newcommand{\BV}{\boldsymbol{B}}
\newcommand{\HV}{\boldsymbol{H}}
\newcommand{\MV}{\boldsymbol{M}}
\newcommand{\be}{\begin{equation}}
\newcommand{\ee}{\end{equation}}
\newcommand{\ba}{\begin{eqnarray}}
\newcommand{\ea}{\end{eqnarray}}
\newcommand{\bq}{\begin{eqnarray*}}
\newcommand{\eq}{\end{eqnarray*}}
\newcommand{\pa}{\partial}
\newcommand{\f}{\frac}
\newcommand{\FV}{\boldsymbol{F}}
\newcommand{\ve}{\boldsymbol{v}}
\newcommand{\AV}{\boldsymbol{A}}
\newcommand{\jv}{\boldsymbol{j}}
\newcommand{\LV}{\boldsymbol{L}}
\newcommand{\SV}{\boldsymbol{S}}
\newcommand{\av}{\boldsymbol{a}}
\newcommand{\qv}{\boldsymbol{q}}
\newcommand{\QV}{\boldsymbol{Q}}
\newcommand{\ev}{\boldsymbol{e}}
\newcommand{\uv}{\boldsymbol{u}}
\newcommand{\KV}{\boldsymbol{K}}
\newcommand{\ro}{\boldsymbol{\rho}}
\newcommand{\si}{\boldsymbol{\sigma}}
\newcommand{\thv}{\boldsymbol{\theta}}
\newcommand{\bv}{\boldsymbol{b}}
\newcommand{\JV}{\boldsymbol{J}}
\newcommand{\nv}{\boldsymbol{n}}
\newcommand{\lv}{\boldsymbol{l}}
\newcommand{\om}{\boldsymbol{\omega}}
\newcommand{\Om}{\boldsymbol{\Omega}}
\newcommand{\Piv}{\boldsymbol{\Pi}}
\newcommand{\UV}{\boldsymbol{U}}
\newcommand{\iv}{\boldsymbol{i}}
\newcommand{\nuv}{\boldsymbol{\nu}}
\newcommand{\muv}{\boldsymbol{\mu}}
\newcommand{\lm}{\boldsymbol{\lambda}}
\newcommand{\Lm}{\boldsymbol{\Lambda}}
\newcommand{\opsi}{\overline{\psi}}
\renewcommand{\tan}{\textrm{tg}}
\renewcommand{\cot}{\textrm{ctg}}
\renewcommand{\sinh}{\textrm{sh}}
\renewcommand{\cosh}{\textrm{ch}}
\renewcommand{\tanh}{\textrm{th}}
\renewcommand{\coth}{\textrm{cth}}

\begin{abstract}
An imperfect cosmic fluid with energy flux is analyzed. Even though its energy density $\rho$ is positive, the pressure $p = -\rho$ due to the fact that the metric is asymptotically de Sitter. The kinematical quantities for a nongeodesic congruence are computed. The scalar expansion is time independent but divergent at the singularity $r = 2m$. Far from the central mass $m$ and for a cosmic time $\bar{t} << H^{-1}$, the heat flux $q$ does not depend on Newton's constant $G$. 

 \end{abstract}
 
\section{Introduction}
 The exterior Schwarzschild solution of Einstein's field equations describes the stationary gravitational field around a spherically symmetric star or black hole (BH). Such asymptotically flat solutions apply to a region outside the central object, far from other compact sources.
 
 A 2nd set of solutions are the cosmological ones that model the behaviour of the universe at its largest scale \cite{CG} as, for example, the homogeneous and isotropic Friedmann - Robertson - Walker (FRW) metric on which the standard model is based. The purpose is to combine both classes of solutions and to find exact solutions for the gravitational field of a compact object embedded in a cosmological background \cite{CG}. One means the solutions describe BHs which are asymptotically flat and, in addition, dynamical \cite{VF, FJ, SHMS, SA, MA}. A complication arises when a BH is ''immersed'' in a cosmological background (excepting the static de Sitter case): the phenomenon of accreation will be present. Namely, a radial flow of energy of the cosmic fluid tends to infall forward the central object \cite{CG, VF, HPI, BN}. 
 
 McVittie \cite{MV}, in his well known paper, has taken into account the effect of the cosmic expansion on local systems. To forbid the accretion of the cosmic fluid into the central object he imposed a special relation between the (time dependent) mass of the object and the scale factor. Since in this case the stress tensor has no non-diagonal components, there is no radiative heat flow. However, the physical meaning of the McVittie spacetime is not completely clear and is still under debate.
 
 Another exact solution that models a spherical object in a cosmological context was proposed by Sultana and Dyer \cite{SD}. Their metric is conformally equivalent to the exterior Schwarzschild metric and the cosmic matter is composed of two non - interacting perfect fluids, one being pressureless dust and the other - a null fluid. In addition, the Sultana - Dyer spacetime does not belong to the class of McVittie models (it is not spatially - Ricci isotropic \cite{CG}). \\
 Throughout the paper we use the geometrical units $G = c = 1$.
 
 \section{Time dependent conformal Schwarzschild metric}
 Let us begin with a general geometry
  \begin{equation}
  ds^{2} = a^{2}(t) \left[-(1- \frac{2m}{r}) dt^{2} + (1- \frac{2m}{r})^{-1} dr^{2} + r^{2} d \Omega^{2}\right], 
 \label{2.1}
 \end{equation}
 where $a(t)$ is a positive function of the time $t$, $m$ is the BH mass (or the mass of a spherically symmetric compact object) and $r > 2m$. When $m = 0$, (2.1) gives us a conformally flat spacetime. In contrast, when $a(t) = 1$ we obtain the exterior Schwarzschild metric around the isolated mass $m$. 
 
 With the help of the GrTensor II package and software package Maple, the following expressions for the components of the Einstein's tensor $G^{a}_{~b}$ are obtained :
 \begin{equation}
 \begin{split}
 G^{0}_{~0} = - \frac{3r\dot{a}^{2}}{a^{4}(r - 2m)},~~~G^{0}_{~1} = - \frac{2m\dot{a}}{a^{3}(r - 2m)^{2}},~~~G^{1}_{~0} = \frac{2m \dot{a}}{a^{3}r^{2}}\\  
 G^{1}_{~1} = G^{2}_{~2} = G^{3}_{~3} = \frac{r(\dot{a}^{2} - 2a \ddot{a})}{a^{4} (r - 2m)},
 \end{split}
\label{2.2}
\end{equation}
 where $\dot{a} = da/dt$, the Latin indices run from $0$ to $3$ and the coordinates are in order $(t, r, \theta, \phi)$. We observe that in (2.2) there are nonzero non - diagonal components. That means a flux of radial energy will be present in the system. Therefore, the cosmic fluid must be imperfect \cite{FJ} and the corresponding energy - momentum tensor given by the Einstein equations $G_{\mu \nu} = 8 \pi T_{\mu \nu}$ should have terms dependent on the heat flux vector. 
 It is worth noting that the scalar curvature
 \begin{equation}
  R^{a}_{~a} = \frac{6 \ddot{a}}{a^{3}(1 - \frac{2m}{r})} 
 \label{2.3}
 \end{equation}
 is singular at $r = 2m$, a property valid also for the McVittie spacetime \cite{FJ}, where the mass $m$ is time dependent. In addition, the metric (2.1) has another (expected) singularity at $r = 0$, where the Kretschmann scalar $R^{abcd}R_{abcd}$ diverges. As far as the Weyl tensor is concerned, all its components are of the form $a^{2}(t)~ C^{Schw}_{abcd}$, where $C^{Schw}_{abcd}$ corresponds to the Schwarzschild metric. The Weyl tensor is, of course, vanishing when $m = 0$ because the metric is conformally flat in that case.
 
 \section{Imperfect fluid stress tensor}
 Let us choose now the following congruence of observers
  \begin{equation}
 u^{a} = \left(\frac{1}{a(t) \sqrt{1 - \frac{2m}{r}}}, 0, 0, 0\right) 
 \label{3.1}
 \end{equation}
with $u^{a}$ the velocity vector field associated to ''static'' observer ($dr/d\tau = 0, u^{a}u_{a} = -1$, $\tau$ being here the proper time). We assume the stress tensor is given by the following expression
  \begin{equation}
  T_{ab} = (p + \rho) u_{a} u_{b} + p g_{ab} + u_{a} q_{b} + u_{b} q_{a},
 \label{3.2}
 \end{equation}
where $u^{a}q_{a} = 0$. The energy density $\rho$, the pressure $p$ and the heat flux $q^{a}$ acquire the form
  \begin{equation}
  \rho = T_{ab}u^{a}u^{b},~~~p = T^{r}_{r} = T^{\theta}_{\theta} = T^{\phi}_{\phi},~~~q^{a} = -T^{a}_{b}u^{b} - \rho u^{a}.
 \label{3.3}
 \end{equation}
Using now Eqs. (2.2), one obtains
\begin{equation}
\begin{split}
\rho = \frac{3 \dot{a}^{2}}{8 \pi a^{4} \left(1 - \frac{2m}{r}\right)},~~~p = \frac{\dot{a}^{2} - 2a \ddot{a}}{8 \pi a^{4} \left(1 - \frac{2m}{r}\right)}, ~~~\\
q^{a} = \left(0,~- \frac{m \dot{a}}{4 \pi a^{4}r^{2}\sqrt{1 - \frac{2m}{r}}},~0,~0\right).
\end{split}
\label{3.4}
\end{equation}
Let us consider the case of an expanding universe ($\dot{a} > 0$). Hence, $q^{1} \equiv q^{r} < 0$, namely a radial inflow will be present. In addition, the parameters $\rho, p$ and $q^{r}$ are divergent at $r = 2m$, a consequence of the singularity of the curvature invariant scalars there. 

 It is worth to compute the total radial energy flowing accross a $r = const.$ hypersurface $\Sigma$ \cite{GH} for some time interval $\Delta t = t_{2} - t_{1}$
   \begin{equation}
   W = \int T^{a}_{~b} u^{b} n_{a} \sqrt{- h}~ dt~ d\theta ~d\phi
 \label{3.5}
 \end{equation}
 where $n^{a} = (0, \sqrt{1 - \frac{2m}{r}}/a, 0, 0)$ is the unit spacelike normal vector to $\Sigma$, $h_{ab} = g_{ab} - n_{a}n_{b}$ is the induced metric on $\Sigma$ and $h = det(h_{ab})$. By means of $T^{1}_{~0} = m\dot{a}/4 \pi a^{3}r^{2}$ from (2.2) and $u^{a}$ from (3.1), we get
 \begin{equation}
  W = \frac{m}{\sqrt{1 - \frac{2m}{r}}} \int^{t_{1}}_{t_{2}} \dot{a}(t)dt = \frac{m\Delta a(t)}{\sqrt{1 - \frac{2m}{r}}}   
 \label{3.6}
 \end{equation}
where $\Delta a(t) = a(t_{2}) - a(t_{1})$. For an observer located at $r >> 2m$, we have $W \cong m\Delta a(t)$, i. e. the gain in the physical mass of the compact object. In other words, all the accreted energy contributes to the increase of the object mass. We of course have neglected here the back reaction effect, keeping $m = const.$ in the metric (2.1). Moreover, $W$ grows indefinitely when $r \rightarrow 2m$. If $\Delta t = t_{2} - t_{1}$ is small, so is $\Delta a$ and, therefore, $W$ is also small excepting near the horizon $r = 2m$.

The congruence (3.1) is not geodesic as could be seen from the expression of the acceleration vector
  \begin{equation}
  a^{b} = \left(0,~\frac{m}{a^{2}r^{2}},~0,~0\right),
 \label{3.7}
 \end{equation}
 with $\sqrt{a^{b} a_{b}} = m/ar^{2}\sqrt{1 - \frac{2m}{r}}$. As far as the other kinematical quantities is concerned, the shear and the vorticity tensors of the congruence are vanishing (due to the spherical symmetry) while the expansion scalar is given by 
   \begin{equation}
  \Theta \equiv  \nabla_{a}u^{a} = \frac{3 \dot{a}}{a^{2} \sqrt{1 - \frac{2m}{r}}} 
 \label{3.8}
 \end{equation}
 If we formally compute the surface gravity on the horizon $r = 2m$, one obtains
   \begin{equation}
   \sqrt{a^{b} a_{b}}~~ \sqrt{- g_{00}}|_{r = 2m} = \frac{m}{(2m)^{2}} = \frac{1}{4m},
 \label{3.9}
 \end{equation}
as for a static BH. It is, however, problematic whether one might define a surface gravity in our nonstatic spacetime and, in addition, at $r = 2m$ the metric is singular. 

\section{Time dependent Schwarzschild - de Sitter geometry}
Our next task is to take into consideration a particular expression of $a(t)$ and analyze that case in detail. A situation not studied till now, to our knowledge, is the Schwarzschild BH embedded in a de Sitter universe, written in conformally flat coordinates. We prefer a spacetime conformal to the Schwarzschild one in order to preserve the null trajectories and the causal horizon.

Let us firstly consider the de Sitter line element written in comoving coordinates
  \begin{equation}
  ds^{2} = -d \bar{t}^{2} + e^{2H\bar{t}} (dr^{2} + r^{2} d\Omega^{2}) , 
 \label{4.1}
 \end{equation}
where $H = 1/\bar{R}$ is the Hubble constant, $\bar{R}$ is the radius of the de Sitter universe (from the static form of the metric, related to the cosmological constant $\Lambda$, with $1/\bar{R}^{2} = \Lambda/3$, $\bar{t}$ is the comoving time and $d\Omega^{2} = d\theta^{2} + sin^{2}\theta d\phi^{2}$ represents the metric on the unit two-sphere. By means of the transformation \cite{PD} 
  \begin{equation}
 \eta = - \frac{1}{H}e^{-H \bar{t}},~~~(\bar{t} > 0, ~~~-\frac{1}{H} < \eta < 0),
 \label{4.2}
 \end{equation}
the metric (4.1) may be put in a conformally flat form, 
  \begin{equation}
  ds^{2} = \frac{1}{H^{2}\eta^{2}} (-d \eta^{2} +  dr^{2} + r^{2} d\Omega^{2}) , 
 \label{4.3}
 \end{equation}
$\eta$ being the conformal time. We intend to study a BH embedded in the above spacetime, namely in the geometry
  \begin{equation}
  ds^{2} = \frac{1}{H^{2}\eta^{2}} \left[- (1 - \frac{2m}{r}) d \eta^{2} +  \frac{dr^{2}}{1 - \frac{2m}{r}} + r^{2} d\Omega^{2}\right] , 
 \label{4.4}
 \end{equation}
 for to find the influence of an expanding universe on local physics. We have now $a(\eta) = 1/H|\eta|$ which leads to 
  \begin{equation}
  \dot{a} = \frac{da}{d\eta} = \frac{1}{H\eta^{2}}, ~~\frac{\dot{a}}{a} = \frac{1}{|\eta|},~~\ddot{a} = - \frac{2}{H\eta^{3}} > 0,~~ 2a\ddot{a} - \dot{a}^{2} = \frac{3}{H^{2}\eta^{4}}
 \label{4.5}
 \end{equation}
 We notice that $\ddot{a} > 0$, that is the spacetime is accelerating, as it is confirmed for our universe. For the above particular value for $a(\eta)$ we have from (2.2) for the components of the stress tensor
 \begin{equation}
 \begin{split}
  8 \pi T^{0}_{~1} = G^{0}_{~1} = \frac{2mH^{2}\eta}{r^{2}(1 - \frac{2m}{r})^{2}} = - \frac{8 \pi T^{1}_{~0}}{(1 - \frac{2m}{r})^{2}} \\
  8 \pi \rho = -8 \pi T^{0}_{~0} = - 8 \pi p = \frac{3H^{2}}{1 - \frac{2m}{r}}
  \end{split}
 \label{4.6}
 \end{equation}
 We see that the equation of state of the fluid is $p = -\rho$, irrespective of the value of $m$. Moreover, they have no a time dependence. For $r >> 2m$ (far from the central mass), $\rho$ acquires the well-known expression $3H^{2}/8 \pi$ for a spatially flat FRW universe. Even though the energy density of the fluid is positive and the weak energy condition holds, the strong energy condition is not satisfied. We have, indeed
   \begin{equation}
   \left(T_{ab} - \frac{1}{2}g_{ab}T^{c}_{c}\right) u^{a}u^{b} = - \rho < 0,
 \label{4.7}
 \end{equation}
a basic property of the dark energy (DE), due to the negative pressure.

The kinematical quantities $a^{b},~\Theta$ and $q^{a}$ become now
   \begin{equation}
   \begin{split}
   a^{b} = \left(0,~\frac{mH^{2}\eta^{2}}{r^{2}},~0,~0\right),~~\sqrt{a^{b}a_{b}} = \frac{mH|\eta|}{r^{2}\sqrt{1 - \frac{2m}{r}}}\\
   \Theta = \frac{3H}{\sqrt{1 - \frac{2m}{r}}},~~q^{a} = \left(0,~- \frac{mH^{3}\eta^{2}}{4 \pi r^{2} \sqrt{1 - \frac{2m}{r}}},~0,~0\right)
   \end{split}
 \label{4.8}
 \end{equation}
 In addition, (4.8) yields
  \begin{equation}
 \dot{\Theta} \equiv u^{a} \nabla_{a}\Theta = 0,~~~q \equiv \sqrt{q^{a}q_{a}} = \frac{mH^{2}|\eta|}{4 \pi r^{2} \left(1 - \frac{2m}{r}\right)}  
 \label{4.9}
 \end{equation}
We see that, in spite of the time dependance of the metric (4.4), $\Theta$ is a function on the radial coordinate only. One means that for an observer sitting at $r = const.$ the congruence has a constant expansion.  It becomes, however, divergent when $r \rightarrow 2m$ because of the singularity.

An interesting property of the spatial vector field $q^{a}$ that represents the current density of heat is its independence of the Newton constant $G$. Far from the constant mass and for the cosmic time $\bar{t} << H^{-1}$ (or from (4.2), $H|\eta| \approx 1$), the expression of the heat flow becomes (we introduce here the fundamental constants $G$ and $c$)
  \begin{equation}
  q = \frac{c^{4}}{G}\frac{Gm}{c^{2}}\frac{H}{4 \pi r^{2}} = \frac{mc^{2}H}{4 \pi r^{2}}.
 \label{4.10}
 \end{equation}
Let us find, as an example, the ingoing heat flux on the Sun surface, using the above approximation. With $M_{S} \approx 2$x$ 10^{33} g,~R_{S} \approx 7$x$10^{10} cm$ and $H \approx 2.3$x$ 10^{-18} s^{-1}$, we get $q_{S} \approx 6.2$x$ 10^{13} ergs/cm^{2}s = 6.88$x$ 10^{-8} g/cm^{2}s$, or a power of $1.4$x$ 10^{44}$ergs/year. 

\section{Misner - Sharp mass}
For to interpret the spacetime (4.4) as a model for an inhomogeneity in a FRW universe (spatially flat in our situation) it is useful to compute the quasi-local Misner - Sharp (MS) mass $M$ \cite{HPFT, NY, GGMP, FMM}, with its Weyl and Ricci parts, namely $M_{W}$ and $M_{R}$, respectively. It will help to detect localized sources of gravity . The mass $M$ is obtained from
  \begin{equation}
  1 - \frac{2M(t,r)}{R} = g^{ab}R_{,a}R_{,b},
 \label{5.1}
 \end{equation}
 where $R_{,a} \equiv \partial R/\partial x_{a}$ is the areal (physical) radius which in our case is given by $R(t,r) = ra(t)$ (see 2.1). Eq. (5.1) yields
   \begin{equation}
   M(t,r) = ma + \frac{\dot{a}^{2} r^{3}}{2a(1 - \frac{2m}{r})}
 \label{5.2}
 \end{equation}
 We identify the 1st term from the r.h.s. with the Weyl mass $M_{W}$ and the 2nd term with the Ricci mass $M_{R}$ that gives the contribution of the fluid to the $MS$ energy. We note that $M_{W}$ is the gravitational energy and $ma(t)$ represents the physical mass. 
 
 Let us now apply the formula (5.2) for the spacetime (4.4). One obtains
   \begin{equation}
   M(\eta,r) = \frac{m}{H|\eta|} + \frac{r^{3}}{2H|\eta| \eta^{2} \left(1 - \frac{2m}{r}\right)} 
 \label{5.3}
 \end{equation}
 The fact that $M_{W} = m/H|\eta|$ grows from $m$ to infinity when $\eta$ varies from $-1/H$ to zero is interpreted as saying that energy is accreated from the ambient matter onto the central object, as one sees from the expression of $q$. $E_{W}$ comes from the Weyl curvature and may be identified with the gravitational mass of the central object. The Ricci part of the Misner - Sharp mass may be expressed in terms of the density $\rho$ from (4.6) and the areal radius, to give $M_{R} = \rho~ 4 \pi R^{3}/3$, as expected (see, for instance, \cite{CG}).

We wish now to evaluate the $MS$ mass at $\bar{t} = 0$ or $|\eta| = 1/H$, for $r >> 2m$. While $M_{W} \approx m$, the Ricci mass appears as 
   \begin{equation}
   M_{R} \approx \frac{H^{2}r^{3}}{2G}
 \label{5.4}
 \end{equation}
where the gravitational constant $G$ has been introduced, to have correct units. Taking again the Sun as the central object and using the previous values of $M_{S}$, $H$ and $R_{S}$, one obtains $M_{R} \approx 2.8$ Kg. We see that by far the basic contribution to $M(\eta,r)$ comes from the Weyl mass. However, the situation changes drastically near $r = 2m$ or for $\eta \rightarrow 0$, when the 2nd term from (5.3) dominates. It is clear that ''to be near the singularity $r = 2m$'' or ''to wait for a long time'' have similar effects.

\section{Conclusions} 
we studied in this paper a BH (or a compact object with spherical symmetry) immersed in a de Sitter space written in a conformally flat form. To be a solution of Einstein's equations, we need an imperfect fluid stress tensor on its RHS. The proposed model leads to the presence of an inward energy flux that diverges at $r = 2m$. 
The kinematical quantities associated to a congruence of non-geodesic observers are calculated. One finds that the shear and vorticity tensors are vanishing but the scalar expansion does not depend on time and is asymptotically constant. The cosmic fluid stress tensor does not observe the strong energy condition  (as for dark energy) due to the negative pressure $p = - \rho$. 
We stress the very simple form of the formula (4.10) for the energy flux ($q$ asymptotically equals the star energy times the Hubble constant divided by its surface area). In addition, it is independent of the gravitational constant.

\end{document}